\documentclass[twocolumn,prc,superscriptaddress]{revtex4}  
\usepackage{xcolor}    
\definecolor{bluelink}{RGB}{0,0,238}

\usepackage{fix-cm}
\usepackage{graphicx}

\usepackage{hyperref}
\hypersetup{
    colorlinks=true,
    linkcolor=blue,
    filecolor=magenta,      
    urlcolor=cyan,
    pdftitle={NFS}
    }

\begin{document}

\title{First beams at Neutrons For Science}



\author{X.~Ledoux}
\author{J.C.~Foy}
\author{J.E.~Ducret}
\author{A.M.~Frelin}
\author{D.~Ramos}
\affiliation{Grand Acc\'el\'erateur National d'Ions Lourds, CEA/DRF - CNRS/IN2P3, B.P.~55027, F-14076 Caen, France}

\author{J.~Mrazek} 
\author{E.~Simeckova}
\author{R.~Behal}
\affiliation{Nuclear Physics Institute of the Czech Academy of Sciences, 250 68, \v{R}e\v{z}, Czech Republic}
\author{L.~Caceres} 
\author{V.~Glagolev}

\affiliation{Nuclear Physics Institute of the Czech Academy of Sciences, 250 68, \v{R}e\v{z}, Czech Republic}
\author{B.~Jacquot} 
\author{A.~Lemasson} 
\author{J.~Pancin} 
\author{J.~Piot} 
\author{C.~Stodel} 
\affiliation{Grand Acc\'el\'erateur National d'Ions Lourds, CEA/DRF - CNRS/IN2P3, B.P.~55027, F-14076 Caen, France}
\author{M.~Vandebrouck} 
\affiliation{IRFU, CEA, Universit\'e Paris-Saclay, 91191 Gif-sur-Yvette, France}




\date{Received: 02/22/2021 / Accepted: date}

\begin{abstract}
The neutrons for science facility (NFS), the first operational experimental area of the new GANIL/SPIRAL-2 facility,
 received its first beams in December 2019. Proton-induced reaction cross-sections as well as neutron beam
 characteristics were measured during the first commissioning phases. The first results, showing the features of the facility, 
are presented here and compared with previously published data.
 The physics cases and the first accepted experiments are presented as well.

\end{abstract}

\maketitle

\section{Introduction}
\label{sec:intro}
GANIL is one of the leading laboratories in the world engaged in research with ion beams \cite{Ganil}. It delivers beams since 1983 for several research topics such as structure of atomic nucleus, nuclear fission, multifragmentation process, hot nuclear matter, nuclear astrophysics or interdisciplinary studies.
The new SPIRAL-2 facility, located on the GANIL site, is composed of a super-conducting linear accelerator and two experimental areas, NFS and S3 \cite{S3}. The accelerator delivers very intense proton, deuteron and heavy ion beams (up to 5~mA). The maximum energy range extends from 14~MeV/A for heavy ions up to 40~MeV for deuterons \cite{Spiral2,Dolegeviez}.

The LINAC is composed of two ion sources (one for protons and deuterons, the other for heavy ions), a radiofrequency quadrupole (RFQ) and 26 super-conducting cavities. This accelerator offered the great opportunity to build the NFS facility. The objective was to produce neutron beams in the 1-40~MeV energy range, unique in the world in terms of flux, energy resolution and background environment. 
This facility will provide new measurements in nuclear physics needs for evaluated data bases as well as for medical applications, detector development, industrial applications and fundamental studies in nuclear physics. \newline
The fission process for instance is not fully understood yet and its study with fast neutrons requires intense beams to deal with low mass samples and small efficiency detectors. Neutron-induced cross-sections measurements on different materials are another example. They are essential for the construction, decommissioning and safety of the nuclear facilities, for nuclear medicine, but also for a deeper understanding of the underlying processes.
The 1-40~MeV energy range is particularly important since it corresponds to the opening of new reaction channels such as (n,xn), (n,n'$\gamma$), or light-charged-particle (LCP) production. In many cases, reaction cross sections are poorly or not known and new measurements are required. 
NFS received its first proton beam on December 6$^{th}$, 2019. For the first time neutrons were generated by the interaction of a 33~MeV proton beam with a Faraday cup. Proton-induced reaction cross-sections were also measured by activation method. After a LINAC shut down period, the first steps of NFS commissioning were performed between September and November 2020. The first neutron beams were produced at NFS by the Be(p,xn) and $^{nat}$Li(p,xn) reactions. Here we report on the first measured neutron spectra and flux and on charged-particle activation measurements. These results are compared with previously published data. A more comprehensive paper will be written on the NFS facility after the full commissioning.

\section{Description}


\subsection{NFS facility}
\label{sec:NFS}
\begin{figure*}
\includegraphics[width=0.95\textwidth]{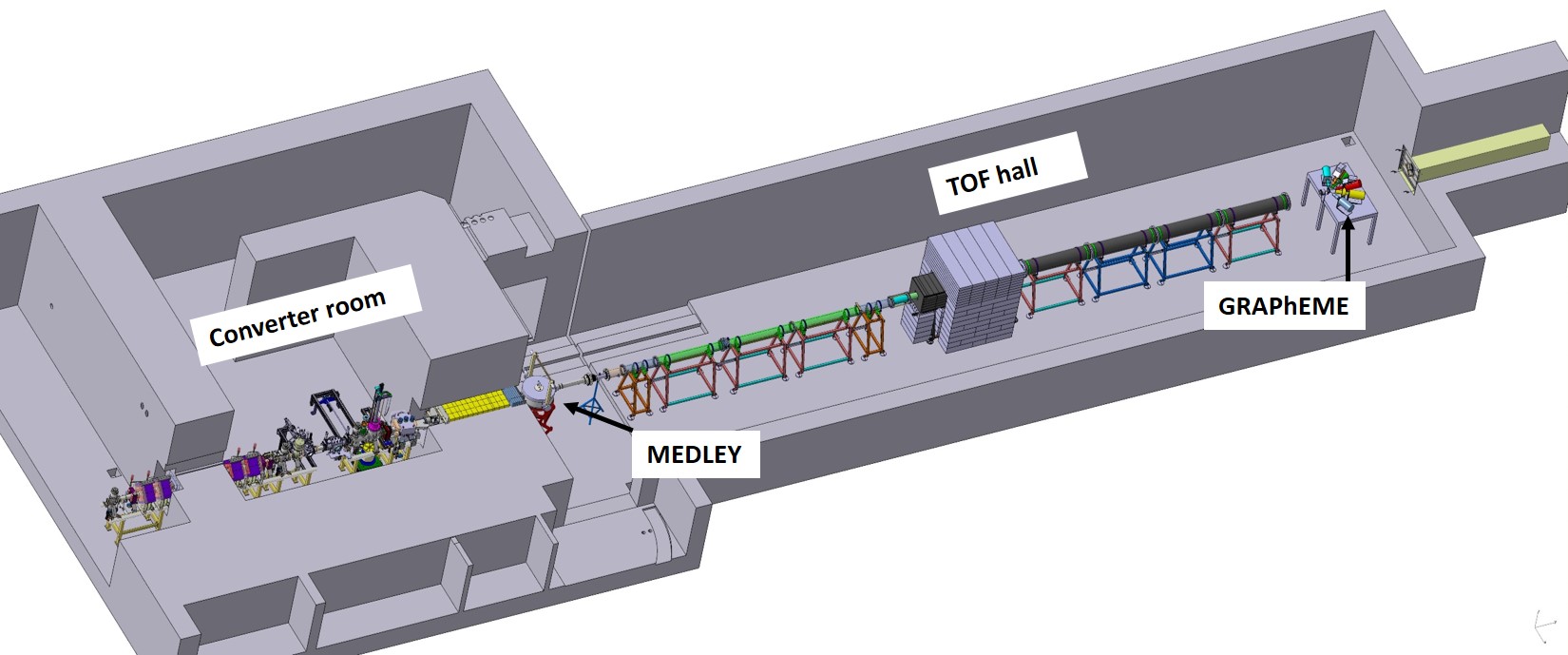}
\caption{Layout of the NFS facility. As an example two detectors that will be used at NFS, MEDLEY \cite{Medley} and GRAPhEME \cite{Grapheme}, are shown.}
\label{shema}       
\end{figure*}

The NFS facility which is at a depth of 9.5~m (like the LINAC) is composed of two main areas: a converter room, where neutrons 
are produced and a time-of-flight (TOF) area (see figure \ref{shema}). The TOF area is an experimental room, 6~m wide and 
28~m long, separated from the converter room by a 3 m thick wall of concrete. A collimator placed inside this wall 
defines the neutron beam in the TOF hall. The collimator, composed of concrete, borated polyethylene, iron and lead, was designed to minimize neutron and gamma background in the TOF area. The internal part of the collimator has a conical shape. The radius of the neutron
 beam spot is estimated to 1.7~cm at the exit of the collimator and to about 13~cm at the end of the TOF hall. The TOF area is equipped with a vacuum beam line to minimize neutron and gamma-ray background generated by the neutron beam. A second collimator is placed in the center of the TOF hall to reduce the neutron beam size. After this collimator the neutron beam spot radius is around 2~cm. The length of the TOF hall allows to carry out several experiments simultaneously at different positions (different flight paths). As the TOF room is equipped with nuclear ventilation, the use of highly radioactive targets, such as actinides, is possible. More details on the NFS facility are available in \cite{Ledoux1,Ledoux2}.

\subsection{Neutron production}
\label{sec:Production}
The interaction of proton or deuteron beams with a lithium or beryllium target is a very efficient way to produce
 energetic neutrons. When a thin target (converter) is used, this production is dominated by the charge-exchange 
  $^7Li(p,n_i)^7Be$, $^9Be(p,n_i)^9B$ reactions where the subscript i denotes the excitation state of the daughter nucleus. 
 Since final states of the reactions are two-body systems, the resulting neutrons have a discrete energy distribution
 defining quasi-mono-energetic neutron beams. Alternatively, the use of a beryllium or carbon converter, thick enough to stop 
 the incident beam, provides neutrons with a continuous energy spectrum. Protons and deuterons primary beams may be used, 
 the latter being more efficient thanks to the deuteron break-up reactions \cite{Saltmarch,Meulders}.
\newline
NFS is equipped with a thick beryllium-rotating converter and two fixed thin converters (lithium and beryllium). These are mounted on a water-cooled target holder
placed at 40.8 cm downstream of the rotating converter. A magnet placed downstream of the fixed converters 
 deflects the primary charged beam toward a Faraday cup in order to measure the beam intensity.
The energy domain of neutrons (0.1 - 40~MeV) and the available flight path (5 to 30~m) require a beam period greater than 1~$\mu$s to avoid the overlap of neutrons from successive bunches.
Hence, the LINAC frequency (88~MHz) must be reduced by a factor of 100 or more. This is obtained by using a single bunch selector {\cite{DiGiacomo} placed downstream of the RFQ to select 1 bunch over N (N$\geq$100). The other bunches are stopped in a beam scraper placed between the RFQ and the LINAC. The maximum intensity sent to NFS is then $\frac{5~mA}{N}$ = 50~$\mu$A (3.12x10$^{14}$ protons or deuterons per second). This also corresponds to the maximum intensity authorized at NFS for the reference reaction, namely 40~MeV d + Be (8~mm). In that case, almost 1.8x10$^{13}$ neutrons are produced per second in 4~$\pi$.sr. When another reaction is used, the maximum intensity is calculated by taking into account radioprotection and thermal constraints and can be greater than 50~$\mu$A.    

\subsection{Irradiation by charged particles}
\label{sec:Irra}
The dedicated experimental setup for charged-particle irradiation was developed in NPI CAS, \v{R}e\v{z} \cite{canam}.
The Irradiation Chamber (IC) \cite{ND2020} is based on an air-lock chamber placed in the converter cave.
The IC is coupled to a pneumatic transport system, developed in KIT Karlsruhe, that transports the samples to/from
the storage in front of a HPGe detector, placed in the TOF area. The IC is equipped with a cooled Faraday
cup for the charge measurements and a cooled revolver-type degrader with twelve positions for rapid beam energy changes.
The time interval between the irradiation end and the positioning of the irradiated sample in front of the HPGe
detector is 45~seconds: 25~seconds for the sample IC extraction and 20~seconds for the transport.
This opens new possibilities to investigate the production of short-lived isotopes.
Sample shuttles, placed in storage (30 positions), can be loaded with a foil stack, where longer-lived isotopes can 
be measured in a range of beam energies, or a single foil, where the beam energy can be lowered by the degrader.

\section{Neutron spectra measurement}
\label{sec:Spectra}
\subsection{Detector}
\label{sec:Detector}
Neutron spectra were measured in the TOF hall with a detector composed of a cylindrical cell ($\phi$=5.08~cm, h=7.62~cm)
 filled with EJ309 liquid scintillator coupled to a phototube. The detector characteristics are detailed in \cite{Fregeau}.
 A pulse shape discrimination (PSD) method using a Mesytec MPD4 module ensured the neutron gamma-ray discrimination.
 The detection threshold was set to 200~keVee, corresponding approximately to a 1~MeV proton energy. The neutron detection 
 efficiency as a function of energy was calculated with the SCINFUL code \cite{Scinful}, its uncertainty being estimated to 5\%. 
 The distance between the converter and the center of the liquid cell was L=1614.9(2)~cm and L=1574.1(2)~cm for the 
 thick and thin converters respectively. The beam diameter at the detector position was estimated to about 12~cm, therefore
larger than the detector. The neutron interaction probability being constant along the cell height (h=7.62~cm), the flight path uncertainty was $\Delta$L=$\frac{h}{2\sqrt3}=2.2~cm$ \cite{BIPM}.

\subsection{Energy measurement}
\label{sec:Energy}
The neutron energy was determined by TOF technique. The beam frequency was set to 22~kHz (N=400). The signal 
 from the scintillation detector was discriminated in a 20\% constant-fraction discriminator module. The resulting output logic 
 signal defined the start and the following LINAC Radio Frequency (RF) signal determined the stop of the TOF in a reverse configuration. 
 In order to cover the TOF energy range (up to 4~$\mu$s), and achieve a time resolution better than 1~ns, start from scintillator 
 and stop (from RF) signals were measured using an ORTEC 566 Time to Amplitude Converter relative to an External Clock with 160~ns period obtained from an ORTEC 462 Time Calibrator. A resulting electronic time resolution of 80~ps was achieved, much better than the bunch duration. 
 The TOF distribution was then transformed into neutron energy distribution using the relationship:
 \label{sec:1}

\begin{equation}
E=\left ( \frac{1}{\sqrt{1-\left (\frac{L}{tc} \right )^2}} \right ) mc^2 
\end{equation}

Considering a time resolution of 1~ns (see next section), a path length L=1574.1~cm and $\Delta$L=2.2~cm, the energy 
resolution for a 30~MeV neutron was better than 0.4~MeV FWHM, namely much better than the proton energy straggling 
in the converters.

\subsection{Yield measurement}
\label{sec:Yield}
The neutron yield at 0 degree is defined as: 

$Y(/MeV/sr/\mu C)=\frac{dN}{dE}\frac{1}{\epsilon \Omega \tau I} $
\newline
with dN/dE the number of neutrons detected by energy bin, $\Omega$ the solid angle covered by the neutron detector, 
$\epsilon$ its efficiency, I the integrated charge of protons in $\mu$C and $\tau$ the acquisition dead 
time correction. For all the spectra presented here, the beam intensity was low enough so that the dead time correction was always smaller than ~2\%.
The polar angle covered by the detector ($\theta\le$ $1.6\times10^{-3}$ rad), being much smaller than the collimator 
aperture ($3.8\times10^{-3}$ rad), defines the solid angle covered by the detector, $\Omega=\pi\frac{(\frac{\phi}{2})^2}{L^2}$. 
With $\Delta$L=2.2~cm, $\frac{\Delta\Omega}{\Omega}$ was estimated to 0.3~\%. 
The charge measurement uncertainty was estimated at 10~\% and 15~\% for thin- and thick- converter runs respectively (see \ref{subsubsec:Thin} and \ref{subsubsec:Thick}).
The resulting total neutron yield uncertainties were 12~\% and 16~\% for thin- and thick- converter runs respectively.

\subsection{Results}

Figure \ref{fig_n_gamma} shows the time distribution between the detector (start) and the LINAC RF signal (stop).
The wide central bump between 400 and 1300~ns corresponds to the neutron distribution. 
The sharp peak on the right hand side corresponds to gamma-rays produced in the converter, equivalent to 
$t_{\gamma}=L/c$. The width of the peak, $\sigma \simeq 1~ns$, reflects the time resolution. This resolution results from
the convolution of the detector time resolution and the bunch width at the converter position. 
The flat background (short time and between neutron bump and gamma peak) corresponds to uncorrelated
events and is subtracted from the time distribution for yield determination. 
A prompt gamma-ray peak is still present in the "neutron identified" events showing that the neutron-gamma rays discrimination is not 
totally effective.

\begin{figure}
 \includegraphics[width=0.45\textwidth]{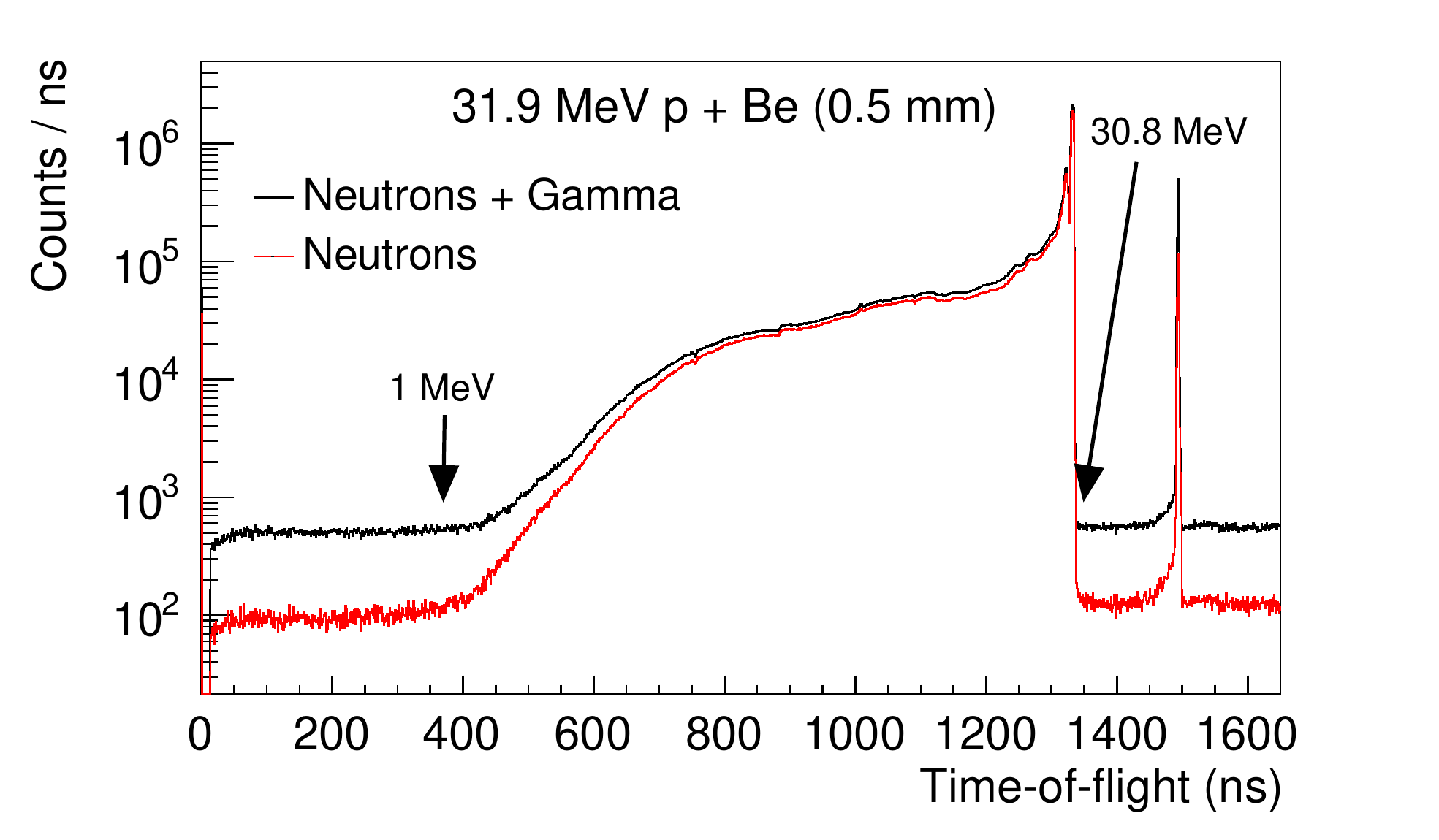}
\caption{Time distributions between the neutron detector and the LINAC RF. Black line corresponds to all events while the red line
 represents the events identified as neutrons by PSD. The arrows give the neutron energy corresponding to the TOF.}
\label{fig_n_gamma}       
\end{figure}

\subsubsection{Thin converter}
\label{subsubsec:Thin}

Two thin converters were used: lithium (1.5~mm thick) and beryllium (0.5~mm thick). 31.9~MeV proton loses approximatively 1.24~MeV
 and 1.37~MeV in Li and Be respectively. Figure \ref{fig_p_thin} shows the measured neutron yields for p+Li and p+Be reactions. 
\newline The neutron spectrum corresponding to p+Li reactions (in black) shows a quasi-mono-energetic peak at 29.7~MeV mainly produced by
 $^7Li(p,n)^7Be$ charge exchange reaction (Q=-1.644~MeV) and a tail at low energy. 
The contributions of $^7Be_g$ and first excited state 
$^7Be_{n1}$ (0.423~MeV) cannot be distinguished as the energy straggling of the protons in the converter is larger than 1.2~MeV. 
The quasi-mono-energetic peak represents approximately half of the neutrons emitted at 0 degree. Our data are compared 
to those published by Uno et al.\cite{Uno}. The improved energy resolution achieved at NFS (in this work) is reflected in the neutron spectra
showing a narrower mono-energetic peak.
\newline  The neutron spectrum corresponding to the p+Be reaction (in red) shows a quasi-mono-energetic peak at 29.5~MeV and a wider distribution centred at 26.9~MeV.
 The peak is mainly produced by $^9Be(p,n)^9B$ charge exchange reaction (Q=-1.850~MeV), while the bump is populated by 
 neutrons from the reactions leaving $^9B$ in the first or second excited state at 2.345~MeV and 2.780~MeV respectively. Theses excited states are not 
separated since their energy separation is much smaller than the proton energy straggling in the converter. Continuum neutrons are also
 produced by other nuclear reactions such as $^9Be(p,pn)^8Be$ and $^9Be(p,n\alpha)^5Li$ among others.
\newline
The p+Li reaction produces more neutrons than the p+Be one (see Table \ref{table11}) and the proportion in the quasi-mono-energetic 
peak is larger (49\% vs 32\% for p+Li and p+Be respectively). The melting points of lithium (180$^{\circ}$C) and berylium (1287$^{\circ}$C) lead to maximum tolerable proton intensities of about 20 and 50~$\mu$A for lithium and beryllium converters respectively. The resulting maximum neutron flux at 5~m from the converter are given in Table \ref{table11}.

\begin{figure}
 \includegraphics[width=0.45\textwidth]{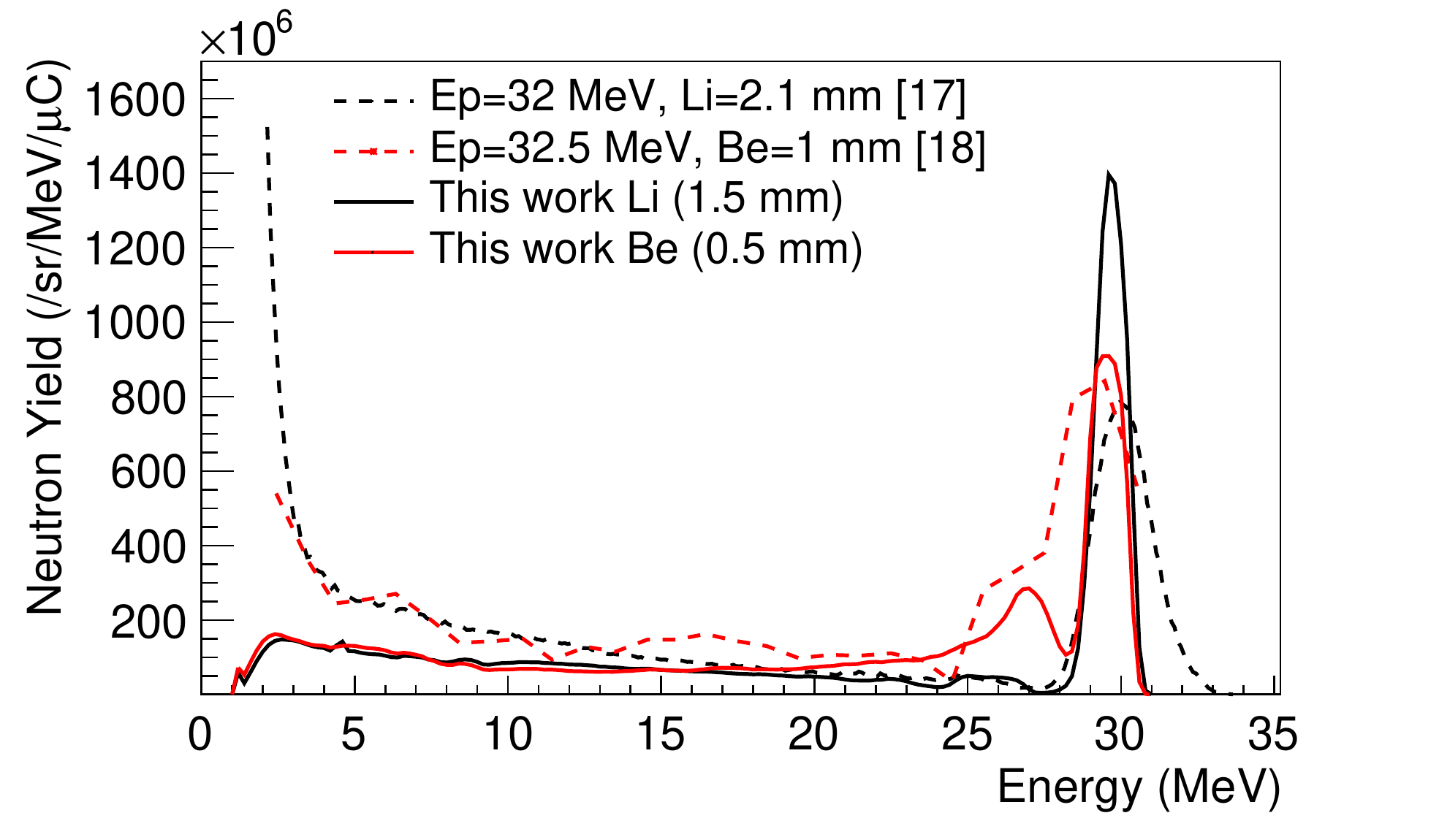}
\caption[width=0.45\textwidth]{Neutron Yield at 0 degree produced by the interaction of 31.9~MeV protons in Be(0.5~mm) and Li(1.5~mm) converters. Solid lines represent this work while dashed lines represent previous publications \cite{Uno,Uwamino}.}
\label{fig_p_thin}       
\end{figure}

\subsubsection{Thick converter}
\label{subsubsec:Thick}

\begin{figure}
 \includegraphics[width=0.45\textwidth]{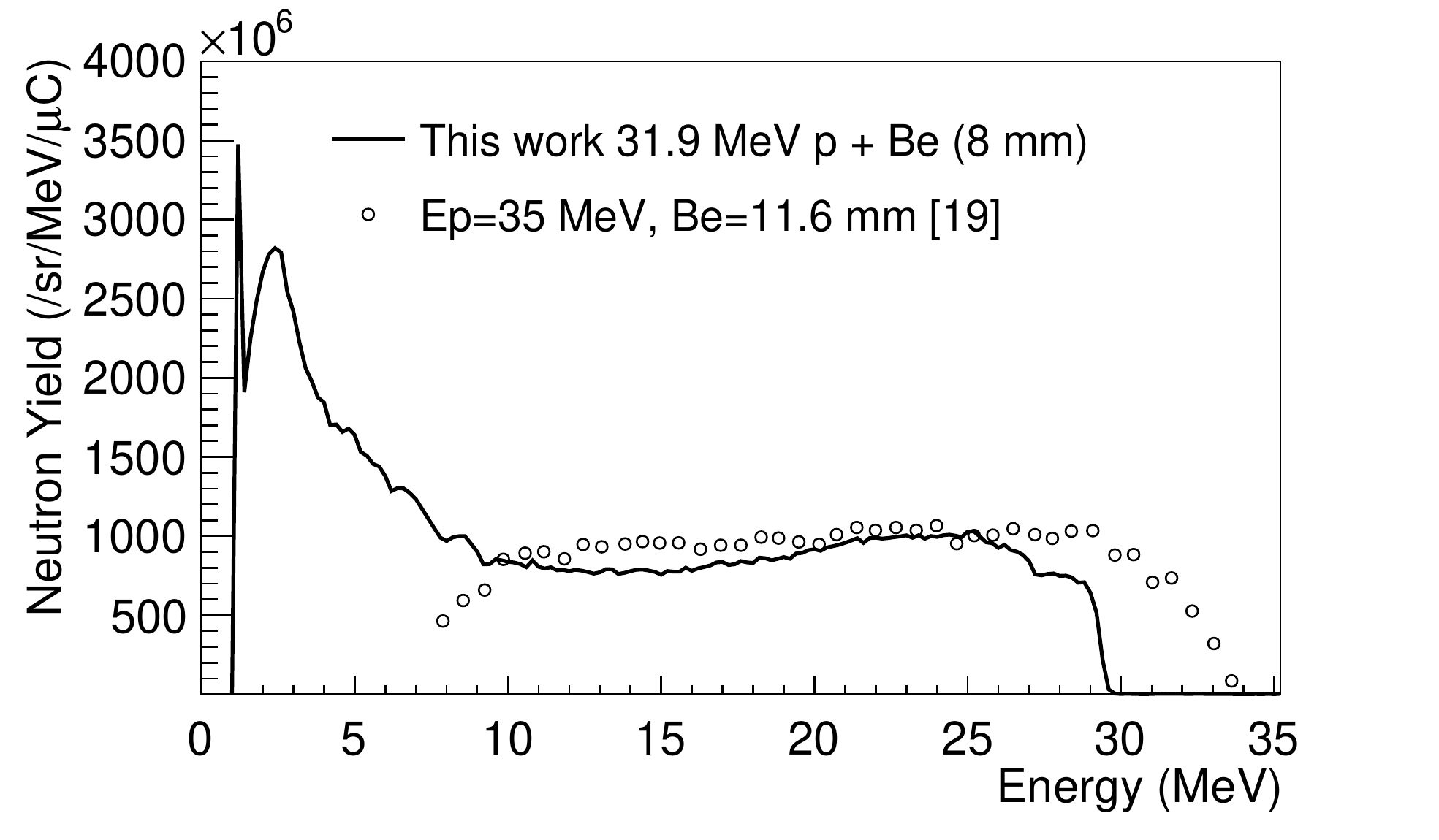}
\caption[width=0.45\textwidth]{Neutron Yield at 0 degree produced by the interaction of 31.9~MeV protons in Be (8~mm). The solid line represent this work while circles represent results from a previous publication \cite{Ullmann}.}
\label{fig_thick}       
\end{figure}

The thick rotating converter of Be (8~mm thick) was irradiated by a 31.9~MeV proton beam stopped in the Be, the proton range being
 6.45~mm. This prevents to measure the primary beam intensity in the Faraday cup. Instead, the beam intensity is therefore measured with the ACCT (Alternative Current Current Transformer) \cite{Jamet}
 of the LINAC placed upstream of the converter. The measured neutron yield is shown in figure \ref{fig_thick} and compared with the 
 data published by Ullmann et al. \cite{Ullmann}. For the latter, the spectrum extends to 33.6~MeV because the incident proton energy was
 higher than in the present measurement. The low-energy part of the spectrum was not covered due to a detection threshold of 7.8~MeV in \cite{Ullmann}.
 The integrated neutron yield is given in Table \ref{table11}. The neutron flux from thick converter is almost one order of magnitude
higher than that from thin converters. This larger flux and the wide energy range illustrate the advantage of using a thick converter.


\begin{table}
\caption{Neutron Yield at 0 degree for E$>$E${}_{min}$, maximum intensity and estimated neutron flux at 5~m.}
\label{table11}       
\begin{tabular}{lllll}
\hline\noalign{\smallskip}
target & $E_{min}$ & Yield (E$>$E${}_{min}$)        & Imax     & Flux at 5~m\\
       & (MeV)     & ($\times10^9n/sr/\mu$C) & ($\mu$A) & ($n.cm^{-2}.s^{-1}$)\\
\noalign{\smallskip}\hline\noalign{\smallskip}
Li (1.5 mm) & 27 & $(1.77 \pm 0.21)$    & 20 & $1.42\times10^5$\\
Be (0.5 mm) & 27 & $(1.61 \pm 0.19)$    & 50 & $3.22\times10^5$\\
Be (8 mm)   & 2  & $(29.4 \pm 4.70)$ & 50 & $5.88\times10^6$\\
\noalign{\smallskip}\hline
\end{tabular}
\end{table}

\section{Proton-induced reaction cross-section measurement}
\label{sec:org9ff6d0d}

\subsection{First experimental test}
\label{sec:orgab157a2}

The IC was developed and tested on the U-120M cyclotron of the Center of Accelerators and
Nuclear Analytical Methods (CANAM) infrastructure of NPI CAS, \v{R}e\v{z} \cite{canam}. In December 2019, LINAC proton beam
tests were conducted at the NFS facility. Four single Cu foils (25~$\mu$m thick) and five single Fe foils (75~$\mu$m thick) were irradiated. 
The proton beam energy was 33~MeV and it was reduced by different degraders.
The foils were irradiated for 1$\textrm{--}$3 minutes with
an intensity of 20$\textrm{--}$80~nA. Each foil was measured after the irradiation for approximately 20 minutes
by the HPGe detector in a event by event regime. After the irradiation session, each sample was measured again
for several hours to obtain data for long-lived isotopes. 

\subsection{Results}
\label{sec:org8b0f3f8}
Fig. \ref{fig_Zn} shows the production cross section of $^{62}$Zn from the monitoring foils obtained in the test at SPIRAL2/NFS (circles)
 and NPI (rectangles). The curve represents the recommended values \cite{Hermanne}.
Fig. \ref{fig_Co} shows example of new data obtained for short-lived $^{54m}$Co with low statistics.
The results show the satisfactory agreement with previous data (Fig. \ref{fig_Zn}) and the ability of the setup to study production cross sections of short-lived isotopes (Fig. \ref{fig_Co}).

\begin{figure}[htbp]
\centering
 \includegraphics[width=0.45\textwidth]{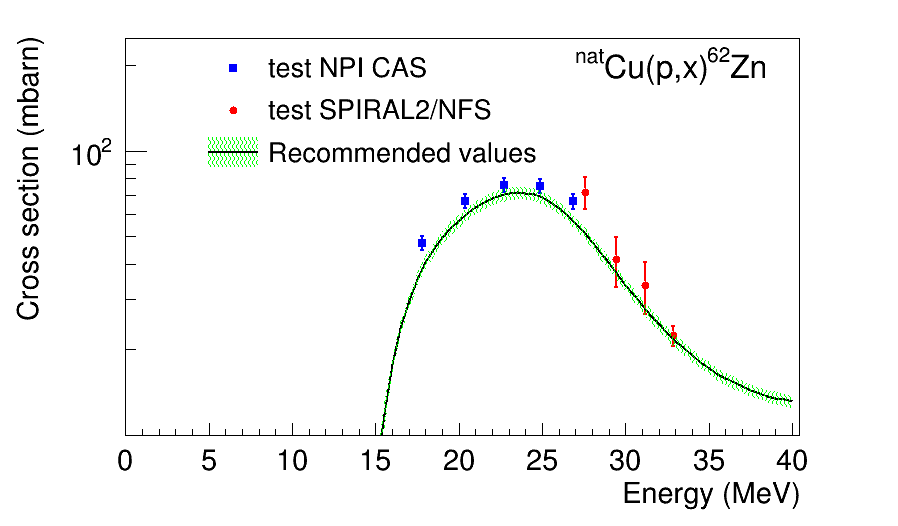}
\caption{Production cross section of $^{62}$Zn from the monitoring foils obtained in the test at SPIRAL2/NFS (red circles) and NPI (blue rectangles). The curve represents the recommended values \cite{Hermanne}.}
\label{fig_Zn}
\end{figure}

\begin{figure}[htbp]
\centering
\includegraphics[width=0.45\textwidth]{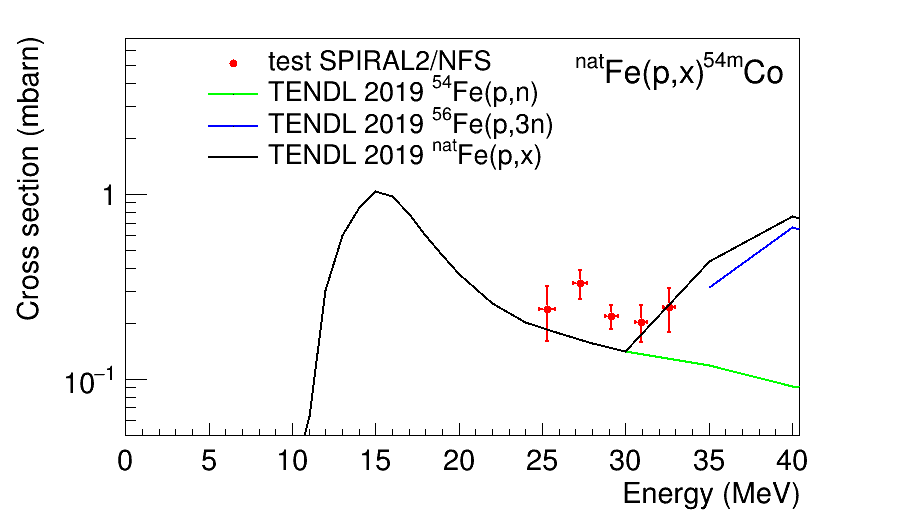}
\caption{Production cross section of $^{54m}$Co (T$_{1/2}$ = 1.48 min) on $^{nat}$Fe target foils, data points obtained during the commissioning. The curves represent model predictions \cite{Koning}.}
\label{fig_Co}
\end{figure}

\section{Discussion and perspectives}
\label{sec:Disc}
The Neutrons For Science commissioning has started and is underway. The measured neutrons fluxes with proton-induced reactions are in good agreement with published data. The neutron energy distributions validated the TOF technique used at NFS for both thin and thick converters and showed very good energy resolution. The commissioning will continue in the next few months with the measurement of neutron beam profiles, neutrons and gamma-ray background. Transmission measurements are also planned in order to investigate the possibilities of total reaction cross-section measurements at NFS. The next step of the commissioning is the use of the 40~MeV deuteron beam at 50~$\mu$A, scheduled for summer 2021, opening up the possibility of new measurements with very low statistical uncertainty. Indeed the d+Be reaction is expected to produce around 10 times more neutrons than the 31.9~MeV p + Be reaction [7]. For the continuous beam the average neutron flux in the TOF hall (short flight path) will be close to $10^8 n.cm^{-2}.s^{-1}$ with a maximum of $5\times10^6 n.cm^{-2}.MeV^{-1}.s^{-1}$ around 14~MeV. Near the converter, the neutron flux will be of the order of $10^{11} n.cm^{-2}.s^{-1}$. NFS facility combines highly desirable characteristics that offer exceptional possibilities
for neutron irradiation with high-energy neutrons.
\newline 
NFS is a new neutron facility with some improvements with respect to other similar facilities such as n-TOF \cite{N-tof}, WNR \cite{WNR} and Gelina \cite{Gelina}. The energy domains, the time structure and the experimental conditions are specific to each installation but between few~MeV and 40~MeV, the average neutron flux will be up to 2 orders of magnitude higher at NFS. 
\newline
To our knowledge, NFS is the only facility in Europe with a collimated quasi-mono-energetic beam in the 10-30~MeV range which offers new perspectives. NFS can, for example, provide reference high-energy quasi-mono-energetic neutron fields from 20~MeV to 30~MeV and extend the range available in the conventional mono-energetic neutron facilities, currently limited to 20~MeV, of the European National Metrological Institutes like PTB, NPL or IRSN \cite{Gressier}. 
\newline
NFS may also benefit from the ion beams accelerated by the LINAC to study reactions induced by ions complementary to those studied at GANIL. 
The installed IC system extends the opportunities to measure activation cross-sections of short-lived isotopes. 
\newline
The first experiments accepted by the GANIL PAC are scheduled for fall 2021. Several topics in neutron- and proton-induced reactions will be covered by these experiments. The study of the production of LCP in neutron-induced reactions, the measurement of O(n,$\alpha$) and $^{238}$U(n,f) reaction cross-section or the study of the fission process are some of the physics cases of these first experiments. All these experiments require a intense neutron flux between few MeV and 40~MeV, a well collimated beam and a low background environment. NFS is the perfect facility for that.

\section{Conclusion}
NFS is expected to achieve commissioning soon, making it an essential installation in the landscape of the neutron facilities. 
Once fully operational, NFS will be an important step forward in the experimental research for both, fundamental and applied research.  
Thanks to its characteristics (high energy, intense flux, energy resolution) new experiments will become feasible: low cross-sections measurement, use of low mass target and/or very low efficiency experimental devices. With NFS, GANIL already attracted new communities of physicists, including those performing nuclear data measurements. Proposals for experiments can already be submitted to the GANIL PAC. NFS is also included in European projects offering Transnational Access such as ARIEL\cite{Ariel} or RADNEXT\cite{Radnext}. 

\begin{acknowledgements}
We warmly thank the engineers and technicians of GANIL and partner laboratories for their valuable contribution to this work. 
This work was supported by MEYS Czech Republic under the projects
SPIRAL2-CZ – OP (\texttt{CZ.02.1.01/0.0/0.0/16\_013/0001679})
and SPIRAL2-CZ (LM2018115).
This work was partially supported by the Seventh Framework Programme through CHANDA (EURATOM contact no. FP7-Fission-2013-605203).
\end{acknowledgements}



\end{document}